\def\be{\begin{equation}}
\def\ee{\end{equation}}
\def\ba{\begin{array}}
\def\ea{\end{array}}

\documentclass[11pt]{article}
\usepackage{amsfonts}
\usepackage{amssymb}
\usepackage{dsfont}
\usepackage{amsmath,amssymb,amstext,amsfonts,array,hhline,tabularx,graphics}
\usepackage{amsthm}

\usepackage{latexsym}
\usepackage{arydshln}
\usepackage{epsfig,graphicx, color}
\usepackage{amsmath}

\newtheorem{thm}{\bf Theorem}

\theoremstyle{plain}
\begin{document}
\parskip=3pt
\parindent=18pt
\baselineskip=20pt \setcounter{page}{1}

 \title{\large\bf Separability Criteria  based on the Weyl Operators}
\date{}

\author{Xiaofen Huang$^{1, 2}$, Tinggui Zhang$^{1, 2}$, Ming-Jing Zhao$^{3}$, Naihuan Jing$^{5, 4}$ \\[10pt]
\footnotesize
\small 1 School of Mathematics and Statistics, Hainan Normal University, Haikou 571158, China\\
\small 2 Key Laboratory of Data Science and Smart Education, \\
\small Ministry of Education, Hainan Normal University,
      Haikou, 571158, China\\
\small 3 School of Science, Beijing Information Science \\
\small        and Technology University, Beijing 100192, China \\
\small 4 Department of Mathematics, Shanghai University, Shanghai 200444, China\\
\small 5 Department of Mathematics, North Carolina State University,                   
Raleigh, NC27695, USA\\}
\date{}

\maketitle

\centerline{$^\ast$ Correspondence to tinggui333@163.com}
\bigskip

\begin{abstract}
 Entanglement as a vital resource for information processing can be described by special properties of the quantum state.
 Using the well-known Weyl basis we propose a new Bloch decomposition of the quantum state and study its separability problem.
 This decomposition enables us to find an alternative characterization of the separability based on the correlation matrix.
 We shaw that the criterion is effective in detecting entanglement for the isotropic states, Bell-diagonal states and some PPT entangled states.
 We also use the Weyl operators to construct an detecting operator for quantum teleportation.

\end{abstract}

\section{Introduction}
Quantum information processing is responsible for implementing
tasks such as super dense coding \cite{code}, teleportation \cite{tele} and key generation \cite{key}. Quantum entanglement is one of the key reasons \cite{chuang} of quantum advantages and has many applications
ranging from quantum teleportation to quantum cryptography \cite{app3}.
In recent years, much effort has been
devoted to understanding entanglement, but still many problems remain unsolved. One key problem is to determine whether a given bipartite state is entangled or separable.
Recall that a bipartite quantum state $\rho$  in a Hilbert space $\mathcal{H}_A\otimes \mathcal{H}_B$ is {\it separable} if
\begin{equation}
\rho=\sum_{i}p_i\rho_{i}^{A}\otimes \rho_i^{B},
\end{equation}where $\{p_i\}$ is a probability distribution,  $\rho_{i}^{A}$  and $ \rho_i^{B}$ are the reduced density matrices of subsystem $\mathcal{H}_A$ and $\mathcal{H}_B$ respectively. Otherwise $\rho$ is said to be {\it entangled}.

 For low dimensional bipartite systems such as $2\otimes 2$, $2\otimes 3$ and $3\otimes 2$ systems, the celebrated PPT criterion \cite{PPT} is a necessary and sufficient condition for separability. However, for higher dimensional multipartite systems, entanglement detection is widely believed to be an NP hard problem. 
 Nevertheless there are several separability criteria available. Among them, a notable one is
 entanglement witness which detects entanglement theoretically and experimentally \cite{entwit}, and most linear separability criterion can be regarded as an entanglement witness. As a nonlinear separability criterion the local uncertainty relation \cite{lur} is an effective method to detect entanglement, and there are some nonlinear criteria based on matrix method, for example, the realignment criterion \cite{realig}, the covariance matrix criterion \cite{cm} and  the separability criterion based on the correlation matrix \cite{sar2020, corr2011}.

In this paper, we focus on an improved Bloch representation of the density matrix
in terms of Weyl basis to derive separability criteria. Our work shows that
 the Weyl basis is advantageous in handling higher dimensional quantum states as well as revealing the symmetry property.
 This new scheme markedly simplifies calculations with density matrices.
Our method further exemplifies Weyl operators can be widely applied in quantum information realm. In fact, Ref. \cite{wapp1} provided the generalized Pauli matrices based on the Weyl operators, and proposed a criterion to detect entanglement by the bounds of the sum of expectation values of any set of anti-commuting observables. Moreover, separability criteria in terms of the Weyl operators for bipartite and multipartite quantum systems were presented
in Ref. \cite{wapp2}. The Weyl operators also play an incredible role in constructing the Weyl discrete channels.
Furthermore, the Weyl operators have been widely used in representation theory of affine Lie algebras and Yangians \cite{liu}.

The layout of the paper is as follows. In section 2 we first show that
Weyl operators provide generalization of the Pauli operators that can be used to represent any quantum state in a tensor format. Based on the
Weyl representation of quantum state, we will propose
a separability criterion in terms of the correlation matrix for an arbitrary bipartite quantum state in section 3.
Our method also gives a necessary and sufficient condition for separability which is applicable in quantum teleportation, as shown in section 4.
Detailed examples are provided to illustrate the advantages of this method compared with previous methods. 

\section{The representation of quantum states in terms of  Weyl operators  }

Let $\mathcal{H}$ be a $d$-dimensional Hilbert space with computational basis $\{|k \rangle\}$, and $\mathds{ Z}_d$ denotes the finite field of modulo $d$ integers. For simplicity all integers in the subscripts are modulo $d$. Recall that
the Weyl operators are defined by
\begin{equation}
W_{nm}:=\sum_{k\in \mathds{ Z}_d}e^{\frac{2 kn\pi {\rm i}}{d}}|k\rangle\langle (k+m) {\rm mod}~ d|,~~ n, m=0, 1, ...,d-1.
\end{equation}
Clearly the set $\{W_{nm}\}$ forms a basis of linear generators in the general linear Lie algebra $\mathfrak{gl}(d)$. We remark that
the Weyl basis is also called the principal basis in the literature. When
$d = 2$, the Weyl operators specialize to the Pauli matrices, i.e., $\{W_{00}, W_{01}, W_{10}, W_{11}\}=\{I, \sigma_1, \sigma_3, {-\rm i}\sigma_2\}$.
In general when $d\geq 2$, the Weyl basis is different from both the Cartan-Weyl and Gell-Mann bases. The Weyl operators $W_{nm}$ enjoy
the following algebraic relations: 
$$
W_{ij}W_{kl}=e^{\frac{2jk\pi{\rm i}}{d}}W_{i+k, j+l}, ~~ W_{kl}^{\dag}= e^{\frac{2kl\pi{\rm i}}{d}}W_{-k, -l}.
$$
Although the Weyl operators are not Hermitian in general, they are unitary and satisfy the orthogonal relation
$$
\langle  W_{nm},  W_{kl}\rangle={\rm { Tr}} W_{nm}^{\dagger}W_{kl}=d\delta_{nk}\delta_{ml},
$$
where $\delta_{ij}=1$ is the Kronecker symbol. Subsequently the Weyl operators obey the trace relation:
\[
{\rm { Tr}} W_{ij}=\left\{
\begin{array}{ccc}
d, &&(i, j)= (0, 0),\\
0, && \text {otherwise.}\\
\end{array}
\right.
\]

As an example, there are  nine linearly independent Weyl operators on a $3$-dimensional Hilbert space listed as follows.
\[
\begin{array}{ccc}
W_{00}=\left(\begin{array}{ccc} 1 & 0 & 0 \\ 0 & 1 & 0 \\ 0 & 0 & 1 \\ \end{array}\right),&
W_{01}=\left(\begin{array}{ccc} 0 & 1 & 0 \\ 0 & 0 & 1 \\ 1 & 0 & 0 \\ \end{array}\right),&
W_{02}=\left(\begin{array}{ccc} 0 & 0 & 1 \\ 1 & 0 & 0 \\ 0 & 1 & 0 \\ \end{array}\right),\\
 W_{10}=\left(\begin{array}{ccc} 1 & 0 & 0 \\ 0 & \omega & 0 \\ 0 & 0 & \omega^2 \\ \end{array}\right),&
 W_{11}=\left(\begin{array}{ccc} 0 & 1 & 0 \\ 0 & 0 & \omega \\ \omega^2 & 0 & 0 \\ \end{array}\right),&
 W_{12}=\left(\begin{array}{ccc} 0 & 0 & 1 \\ \omega & 0 & 0 \\ 0 & \omega^2 & 0 \\ \end{array}\right),\\
 W_{20}=\left(\begin{array}{ccc} 1 & 0 & 0 \\ 0 & \omega^2 & 0 \\ 0 & 0 & \omega \\ \end{array}\right),&
W_{21}=\left(\begin{array}{ccc} 0 & 1 & 0 \\ 0 & 0 & \omega^2 \\ \omega & 0 & 0 \\ \end{array}\right),&
   W_{22}=\left(\begin{array}{ccc} 0 & 0 & 1 \\ \omega^2 & 0 & 0 \\ 0 & \omega & 0 \\ \end{array}\right),
\end{array}
\]
where $\omega$ is the 3th primitive unit root of 1, i.e., $\omega^3=1$.

Since the $d^2$ linearly independent Weyl operators $W_{nm}$ form a basis of $\mathfrak{gl}(d)$, every $d\times d$ density matrix $\rho$ can be uniquely expressed as a linear combination of the Weyl basis:
\begin{equation}\label{Bloch vector}
\rho=\frac{1}{d}(I+\sum_{(i, j)\neq (0, 0)}a_{ij}W_{ij}),
\end{equation}
where the coefficients $a_{ij}={\rm{Tr}} W_{ij}^{\dag}\rho$ for $i, j=0, 1, ..., d-1$.
Since $\rho^\dagger=\rho$, the coefficients satisfy the symmetry condition
\begin{equation}\label{e:sym}
a_{nm}^{\ast}=e^{\frac{-2 nm\pi \rm{i}}{d}}a_{-n, -m},
\end{equation}
where $\ast$ means the complex conjugation.
We also call $\mathbf{\nu}=({a_{ij}})$ as the Bloch vector of $\rho$ relative to the Weyl basis and its length is defined as $|\mathbf{\nu}|=\sum_{i,j} |a_{ij}|^2$.
Therefore any density matrix $\rho$ in Hilbert space $\mathcal{H}$ can be uniquely characterized by a $d^2-1$ dimensional vector $\nu \in \mathbb{C}^{d^2-1}$
with the symmetry condition \eqref{e:sym}, that is, $d^2-1$ real parameters.

\begin{thm}\label{length}
For any $d$-dimensional quantum state $\rho$ in the form of Eq. (\ref{Bloch vector}), the length of the vector $\mathbf{\nu}=\nu(\rho)$ satisfies the following inequality
\begin{equation}
|\mathbf{\nu}|\leq \sqrt{d-1}.
\end{equation}
In particular, the equality holds if and only if $\rho$ is pure. 
\end{thm}

\textbf{Proof:}
Since any quantum state $\rho$  satisfies the trace condition ${\rm Tr}\rho^2\leq 1$, one obtains that
\begin{eqnarray}
{\rm Tr}\rho^2&=&\frac{1}{d^2}{\rm Tr}(I+\sum a_{ij}W_{ij})(I+\sum a_{kl}W_{kl}) \notag \\
&=&\frac{1}{d}+ \frac{1}{d^2}\sum e^{\frac{2jk\pi{\rm i}}{d}}a_{ij}a_{kl}{\rm Tr}W_{i+k, j+l}\leq 1.  \notag
\end{eqnarray}
 Note that $W_{00}=I$ and the matrices $W_{ij}$ are traceless for $(i, j)\neq (0, 0)$, the only nonzero terms in the summation are for $k=-i$ and  $l=-j$, that is,
\begin{equation}
{\rm Tr}\rho^2=\frac{1}{d}(1+ \sum_{(i, j)\neq (0, 0)} e^{\frac{-2ij\pi{\rm i}}{d}}a_{ij}a_{-i, -j})\leq 1.
\end{equation}
Using the symmetry condition \eqref{e:sym}, the trace is simplified as
\begin{equation}
{\rm Tr}\rho^2=\frac{1}{d}(1+\sum_{(i, j)\neq (0, 0)} a_{ij}a_{ij}^{\ast})=\frac{1}{d}(1+|\nu|^2)\leq 1.
\end{equation}
Therefore $|\mathbf{\nu}|\leq \sqrt{d-1}$.
If $\rho$ is pure, we have ${\rm Tr}\rho^2=1$, which means $|\mathbf{\nu}| = \sqrt{d-1}$.
This completes the proof.

Theorem \ref{length} tells us that all Bloch vectors lie within a hypersphere of radius $\sqrt{d-1}$ with the pure states on the spherical surface. Moreover, the quantum state $\rho$ is determined by the Bloch vector $\nu=\nu(\rho)$ satisfying the symmetry condition
\eqref{e:sym}, thus the set of the Bloch vectors is a subset of the vector space $\mathbb{C}^{d^2-1}$ with $d^2-1$ real parameters, i.e. of the size $\mathbb R^{d^2-1}$.

\section{Application of Weyl operators in separability}

With the Weyl basis, a quantum state $\rho$ in space $\mathcal{H}_A\otimes \mathcal{H}_B$ with $dim\mathcal{H}_A=d_A$ and $dim\mathcal{H}_B=d_B$ can be decomposed as
\begin{equation}\label{state}
 \rho=\frac{1}{d_Ad_B}(I_A\otimes I_B+\sum_{(i,j)\neq (0,0)}\alpha_{ij}W_{ij}^{A}\otimes I_B+\sum_{(k,l)\neq (0,0)} \beta_{kl}I_A\otimes W_{kl}^{B}+\sum_{(i,j),(k,l)\neq (0,0)}\lambda_{i,j}^{k,l}W_{ij}^{A}\otimes W_{kl}^{B}),
\end{equation}
where the coefficients $\alpha_{ij}={\rm Tr} \rho(W_{ij}^{A})^{\dag}\otimes I_B$, $\beta_{kl}={\rm Tr} \rho I_A\otimes (W_{kl}^{B})^{\dag} $, $\lambda_{i,j}^{k,l}={\rm Tr} \rho (W_{ij}^{A})^{\dag}\otimes (W_{kl}^{B})^{\dag} $, $I_{A(B)}$ and $W_{ij}^{A(B)}$ are the identity operator and the Weyl operators of the space $\mathcal{H}_{A(B)}$ respectively.
Let $\alpha$ and $\beta$ be two complex vectors of dimension $d_A^2-1$ and $d_B^2-1$ respectively, that is,
\[
\begin{array}{c}
  \alpha=(\alpha_{ij})=(\alpha_{01}, \ldots, \alpha_{0, d_A-1}, \ldots, \alpha_{d_A-1,0 }, \ldots, \alpha_{d_A-1, d_A-1})^t, \\
  \beta=(\beta_{kl})=(\beta_{01}, \ldots, \beta_{0, d_B-1}, \ldots , \beta_{d_B-1, 0}, \ldots, \beta_{d_B-1, d_B-1})^t,
\end{array}
\]
where $t$ denotes transposition.
The entries $\lambda_{i,j}^{k,l}$ form a  matrix $M$ with size
$(d_A^2-1)\times (d_B^2-1)$, which will be referred to as the {\it correlation matrix} of $\rho$ (relative to the Weyl basis).
Since $\rho^\dagger=\rho$, the entries satisfy the symmetry condition: $\alpha_{ij}^{\ast}=e^{\frac{-2 ij\pi \rm{i}}{d_A}}\alpha_{-i, -j}$, $\beta_{kl}^{\ast}=e^{\frac{-2 kl\pi \rm{i}}{d_B}}\beta_{-k, -l}$, {$(\lambda_{i,j}^{k,l})^{\ast}=e^{-2\pi {\rm{i}}( \frac{ij}{d_A}+\frac{kl}{d_B})}\lambda_{-i, -j}^{-k, -l}$}.

According to the decomposition in Eq. (\ref{state}), the reduced states of $\rho$ on the two subsystems are respectively given by
\begin{equation}\label{partial trace A}
  \rho_{A}={\rm Tr}_B \rho=\frac{1}{d_A}(I_A+\sum_{(i, j)\neq(0,0)}\alpha_{ij}W_{ij}^{A}),
 \end{equation}
 \begin{equation}\label{partial trace B}
  \rho_{B}={\rm Tr}_A \rho=\frac{1}{d_B}(I_B+\sum_{(k, l)\neq(0,0)}\beta_{kl}W_{kl}^{B}).
\end{equation}

 \begin{thm}\label{thm2}
A bipartite pure quantum state $\rho=|\psi\rangle\langle\psi|$ in the form of Eq. (\ref{state}) is a product state, i.e., $\rho=\rho_A \otimes \rho_B$, if and only if the correlation matrix $M$ is of rank 1: the matrix $M$ can be written as
 \begin{equation}\label{pure sep}
   M=\alpha \beta^t.
 \end{equation}
 for some column vectors $\alpha$ and $\beta$.
  \end{thm}

{\textbf{Proof:}}
One notices that Eq. (\ref{state}) can be rewritten as
\begin{equation}\label{them2}
\rho=\rho_A\otimes \rho_B+\frac{1}{d_Ad_B}\sum_{(i,j),(k,l)\neq (0,0)}(\lambda_{i,j}^{k,l}-\alpha_{ij}\beta_{kl})W_{ij}^{A}\otimes W_{kl}^{B}.
\end{equation}
Since the matrices $W_{ij}^{A}\otimes W_{kl}^{B}$ are linearly independent, $(\lambda_{i,j}^{k,l}-\alpha_{ij}\beta_{kl})W_{ij}^{A}\otimes W_{kl}^{B}=0$ if and only if $\lambda_{i,j}^{k,l}-\alpha_{i,j}\beta_{k,l}=0$ for $(ij), (kl)\neq (0,0)$, that is, $M=\alpha \beta^t$, and this completes the proof.

 Since any mixed state is a convex combination of pure states, Theorem \ref{thm2} provides a necessary condition for separability for any mixed state in a bipartite system.

Now we denote the Ky Fan matrix norm of $M$ as
$\|M\|_{KF}=\sum \xi_i=\rm{Tr}\sqrt{M^\dag M}$, which is the sum of the singular values $\xi_i$ of the matrix $M$. Then we have the following necessary condition for separability for any bipartite quantum state.
\begin{thm}\label{th5}
 If a bipartite state $\rho$ in the form of Eq. (\ref{state}) is separable, then it has
\begin{equation}\label{sep}
\|M\|_{KF}\leq\sqrt{(d_A-1)(d_B-1)}.
\end{equation}
\end{thm}

\textbf{Proof:}
Suppose the quantum state $\rho$ is separable, then there exist a series of $p_s$, $\rho^{A}_s$, $\rho^{B}_s$ such that $\rho=\sum_s p_s \rho^{A}_s\otimes \rho^{B}_s$, with $p_s\geq 0$ and $\sum_s p_s=1$. Suppose $\rho_s^{A}=\frac{1}{d_A}(I_A+\sum_{(i, j)\neq(0,0)}\alpha_{ij}^{(s)} W_{ij}^{A})$ and $ \rho_s^{B}=\frac{1}{d_B}(I_B+\sum_{(k, l)\neq(0,0)}\beta_{kl}^{(s)} W_{kl}^{B})$ with Bloch vectors $\alpha_{s}=(\alpha_{ij}^{(s)})$ and $\beta_{s}=(\beta_{ij}^{(s)})$, respectively.
Therefore the correlation matrix $M$ of $\rho$ is $M=\sum_s p_s \alpha_{s} \beta_{s}^t$. One sees that
\begin{eqnarray}
\|M\|_{KF} &\leq &\sum_{s} p_s \|\alpha_{s} \beta_{s}^t \|_{KF} \leq \sum_{s} p_s |\alpha_{s}| |\beta_{s} |  \notag  \\
   &\leq & \sum_{s}p_s\cdot\sqrt{d_A-1}\cdot\sqrt{d_B-1}=\sqrt{(d_A-1)(d_B-1)}.
\end{eqnarray}
This completes the proof.

\textbf{Example 1}
Consider the isotropic state $\rho_{iso}=(\frac{1-p}{d^2})I\otimes I+p |\psi_{+}\rangle\langle\psi_{+}|$, where $0 \leq p\leq 1$. They are separable if and only if $p\leq \frac{1}{d+1}$ \cite{iso}.  Since the maximally entangled pure state $|\psi _{+}\rangle=\frac{1}{\sqrt{d}}\sum_{i=0}^{d-1}|ii\rangle$ has the Bloch decomposition relative to Weyl operators:
\begin{equation}\label{rest2}
|\psi_{+}\rangle\langle\psi_{+}|=\frac{1}{d^2}I\otimes I+\sum_{(i,j)\neq (0,0)} \frac{1}{d^2}W_{ij}\otimes  W_{-i, -j}.
\end{equation}
Therefore, the isotropic state can be represented as
\begin{equation}
\rho_{iso}=\frac{1}{d^2}(I\otimes I+\sum_{(i,j)\neq (0,0)}pW_{ij}\otimes W_{-i, -j}).
\end{equation}
Note that the Ky Fan norm  of the correlation matrix $M$ of $\rho_{iso}$ is $\|M\|_{KF} =(d^2-1)p$. Theorem \ref{th5} implies that $p\leq \frac{1}{d+1}$ when the isotropic state is separable.  This means that we can detect all entangled isotropic states by Theorem \ref{th5}.

\textbf{Example 2} Let $\rho$ be
the following $3\times 3$ PPT entangled state found in \cite{ch}:
\begin{equation}\label{e3}
\rho=\frac{1}{4}(I-\sum_{i=0}^4|\chi_i\rangle\langle\chi_i|),
\end{equation}
where $|\chi_0\rangle=|0\rangle(|0\rangle-|1\rangle)/\sqrt{2}$, $|\chi_1\rangle=(|0\rangle-|1\rangle)|2\rangle/\sqrt{2}$, $|\chi_2\rangle=|2\rangle(|1\rangle-|2\rangle)/\sqrt{2}$, $|\chi_3\rangle=(|1\rangle-|2\rangle)|0\rangle/\sqrt{2}$, $|\chi_4\rangle=(|0\rangle+|1\rangle+|2\rangle)(|0\rangle+|1\rangle+|2\rangle)/3$.
We get the Ky Fan norm of the correlation matrix $\|M\|_{KF}$  approximately equals to 2.15, which violates the inequality in Theorem \ref{th5}. Therefore the state $\rho$ is entangled.

\textbf{Example 3}
The Bell-diagonal states can be represented as $\rho=\frac{1}{4}(I+\sum_{i=1}^3 t_i\sigma_i \otimes \sigma_i)$, where $\sigma_i$ are the Pauli operators \cite{M.Zhao2015}. The Bell-diagonal states are known to be separable iff $|t_1|+|t_2|+|t_3|\leq 1$ \cite{M.Zhao2015}.
Consider the Bloch decomposition of $\rho$ relative to the Weyl operators:
\begin{equation}
\rho=\frac{1}{4}(I\otimes I+t_1 W_{01}\otimes W_{01}+t_3W_{10}\otimes W_{10}-t_2W_{11}\otimes W_{11}).
\end{equation}
Then the Ky Fan norm of the correlation matrix is $\|M\|_{KF}=|t_1|+|t_2|+|t_3|$. It follows from Theorem \ref{th5}
that $|t_1|+|t_2|+|t_3|\leq 1$ when the Bell-diagonal states are separable. Again Theorem \ref{th5} completely detects the entanglement for all Bell-diagonal states.

\section{Application of Weyl operators in 
quantum teleportation}

In the process of quantum teleportation, the optimal fidelity of
teleportation as an entangled resource can be expressed by the fully entangled fraction \cite{f1, f2, f3}.
For a given quantum state $\rho$ in a $d$-dimensional Hilbert space, the optimal fidelity of teleportation with respective to  $\rho$ can be described by the
function
\begin{equation}\label{fi}
  f_{max}(\rho)=\frac{dF(\rho)}{d+1}+\frac{1}{d+1},
\end{equation}
where $F(\rho)$ is the fully entangled fraction with respect to $\rho$ defined by \cite{f2}:
\begin{equation}\label{fr}
  F(\rho)=\max_{U}\langle\psi _{+}|(U^{\dagger}\otimes I)\rho (U\otimes I)|\psi _{+}\rangle,
\end{equation}
where $U$ runs through all $d\times d$ unitary matrices, $I$ is the $d\times d$  identity matrix, and $|\psi _{+}\rangle$ is the maximally entangled state.
A state $\rho$ is a useful resource for teleportation if and only if $F(\rho)> \frac{1}{d}$ \cite{f2}. If $F(\rho)\leq\frac{1}{d}$, the fidelity is considered to be not better than separability.
In this sense, the fully entangled fraction $F(\rho)$ can be used to detect quantum
teleportation resource. Ref. \cite{efe1} gave an elegant formula for a two qubit system  by using the method
of Lagrange multipliers. Refs. \cite{N. Ganguly, M. Zhao2012} constructed the teleportation witness for detecting the quantum states that are useful for quantum teleportation.

Now we construct an operator using the Weyl representation to detect if a quantum state is useful for quantum teleportation.
Since the maximally entangled state $|\psi _{+}\rangle=\sum_{i} \frac{1}{\sqrt{d}}|ii\rangle$ can be decomposed as Eq. (\ref{rest2})
according to the Weyl operators \cite{huang},
we let $P_{ij}=UW_{ij}U^{\dag}_U$ and define a normal operator $\mathcal{O}$ by
\begin{eqnarray}\label{o1}
  \mathcal{O}_U :=I\otimes I+ \sum _{(i, j)\neq (0, 0)} P_{ij}\otimes W_{-i, -j}.
\end{eqnarray}
We claim that the operator $\mathcal{O}_U$  can be used to detect whether an unknown quantum state is available for  quantum teleportation.

\begin{thm} The quantum state $\rho$ in a quantum system $\mathcal{H}\otimes \mathcal{H}$ with $dim(\mathcal{H})=d$ is useful for teleportation if and only if there exists some unitary operator $U$ such that the mean value 
satisfies the inequality:
\begin{equation}
\langle\mathcal{O}_U\rangle_{\rho}> d.
\end{equation}
\end{thm}

\textbf{Proof.} For any quantum state $\rho$, one has
\begin{eqnarray*}
\langle\mathcal{O}_U \rangle_{\rho}&=&\langle I\otimes I+ \sum _{(i, j)\neq (0, 0)} P_{ij}\otimes W_{-i, -j}\rangle_{\rho}\\
&=&\langle I\otimes I+\sum _{(i, j)\neq (0, 0)} UW_{ij}U^{\dag}\otimes W_{-i, -j}\rangle_{\rho}\\
&=& d^2 \langle U\otimes I|\psi _{+}\rangle \langle \psi _{+}|U^{\dag}\otimes I\rangle_{\rho}
\end{eqnarray*}

Since $\max_{U} \langle\mathcal{O}_U\rangle_{\rho}=d^2F(\rho)$, and quantum state is useful for quantum teleportation if and only if $F(\rho)> \frac{1}{d}$.
Note that the maximum value is attainable since $\rm{SU}(d)$ is compact. Therefore the quantum state is useful for quantum teleportation if and only if there exists a unitary operator $U$ such that $\langle\mathcal{O}_U\rangle_{\rho}> d$. This completes the proof.

\textbf{Example 4}
Consider the following bipartite state \cite{julio}
\begin{equation}\label{e1}
\rho=p|\phi^{-}\rangle\langle \phi^{-}|+(1-p)|00\rangle\langle00|,
\end{equation}
where $p\in [0,1]$ and
$
|\phi^{-}\rangle=\frac{1}{\sqrt{2}}(|01\rangle- |10\rangle).
$
The PPT criterion establishes that state (\ref{e1}) is separable iff $p=0$ \cite{PPT}.
We choose the operator
$U=|0\rangle\langle 1|+|1\rangle \langle 0|=\sigma_1$,
then one has $\langle\mathcal{O}\rangle_{\rho}=3p$.
Therefore the quantum state $\rho$ in Eq. (\ref{e1}) is useful for quantum teleportation when $p>\frac{2}{3}$.

\section{Conclusions}
 We have investigated the Bloch decomposition of a density matrix relative to the Weyl basis in the quantum system. The geometric properties of Bloch vectors including the length are described in detail. For bipartite quantum states, we have provided a necessary condition of separability in terms of the Ky Fan norm of the correlation matrix. Furthermore, we have demonstrated feasibility and effectiveness of the separability criterion in detecting entanglement using examples of isotropic states, Bell-diagonal states and some PPT entangled states. Finally, we have constructed an operator based on the Weyl operators for detecting useful resource for quantum teleportation .

\bigskip
\noindent{\bf Acknowledgments}

The research is partially supported by the National Natural Science Foundation of China under Grant Nos. 12126351, 12126314, 11861031, 12171044, 12061029. This project is also supported by the specific research fund of the Innovation Platform for Academicians of Hainan Province under Grant No. YSPTZX202215 and Hainan Provincial Natural Science Foundation of China under Grant No. 121RC539.

\end{document}